# Real-time reconstruction of complex non-equilibrium quantum dynamics of matter

**Authors:** V. Stooß[1], S. M. Cavaletto[1], A. Blättermann[1], P. Birk[1], C. H. Keitel[1], C. Ott[1] and T.Pfeifer[1*]

**Affiliation:**

[1]Max-Planck-Institut für Kernphysik, Saupfercheckweg 1, 69117 Heidelberg, Germany, EU

*Correspondence to: Thomas Pfeifer, email address: thomas.pfeifer@mpi-hd.mpg.de

**One Sentence Summary:** A single spectrum of ultrashort signals transmitted through materials contains its entire holographic (amplitude and phase) time-resolved information, even for the case of nonlinearly driven systems with complex time dependence.

**Causality implies that by measuring an absorption spectrum, the time-dependent *linear* response function can be retrieved [1, 2, 3]. Recent experiments [4, 5, 6, 7] suggest a link between the shape of spectral lines observed in absorption spectroscopy with the amplitude and phase of the system's response function. This has even been observed in the presence of *strong, nonlinear* interactions, which promote the observed system out of equilibrium, making it explicitly time dependent [8, 9, 10]. Thus far, however, only the special case of a sudden modification of the response function was understood analytically, leaving the general case of the *dynamical* response to arbitrary interactions open to interpretation [11]. Here, we demonstrate that even for the case of a strongly driven, time-dependent system, one can reconstruct the *full* temporal response information from a single spectrum if a sufficiently short signal is used to trigger the absorption process. This finding is directly applied to a time-domain observation of Rabi cycling [10] between doubly-excited atomic states in the few-femtosecond regime. This general approach unlocks single-shot real-time-resolved signal reconstruction across time scales down to attoseconds for non-equilibrium states of matter. In contrast to available pump–probe schemes, there is no need for scanning time delays in order to access real-time information. The scientific applications of this technique range from testing fundamental quantum dynamics, to measuring and controlling ultrafast, chemical and biological reaction processes.**

Measuring the fastest dynamical processes in nature relies on observing the *nonlinear* response of a system to precisely timed interactions with external stimuli. [12,13]. This typically requires two (or more) controlled events, e.g., a triggering pump and a delayed probe pulse. To access explicitly time-dependent, excited-state dynamics, attosecond time-resolved absorption spectroscopy [14, 15, 16] was used with great success to uncover the non-equilibrium electron dynamics of atomic [17, 18, 19], molecular [20, 21, 22] and condensed-phase [23, 24, 25] systems in the presence of additional interactions. Such out-of-equilibrium processes include, e.g. the coupling of multi-electron configuration channels [8], strong-field manipulation of autoionization [9], or the strong coupling between excited quantum states [10]. Accessing and understanding the non-linear response of such processes is of crucial importance for controlling and steering quantum dynamics on the attosecond time scale [26, 27, 28, 29].

In the *linear* regime, time-domain information can be directly obtained from spectroscopic data. The reason is, that causality connects amplitude (absorption) and phase (dispersion) by means of the Kramers–Kronig relations [1,2,3]. For systems in a stationary state the linear absorption spectrum $A(\omega)$ or cross section $\sigma(\omega)$ is proportional to the Fourier transform of the response function $d(t)$ caused by a probe field $\mathcal{E}_\gamma(t)$:

$$A(\omega) \propto \sigma(\omega) \propto \omega\, Im\left\{\frac{\mathcal{F}[d(t)]}{\mathcal{F}[\mathcal{E}_\gamma(t)]}\right\} \quad \text{eq.(1)}.$$

Thus, the system's response carries information about its internal structure (e.g. resonant excitations) including the natural decay dynamics in characteristic time scales $T_{nat}$. For all-linear interactions, the decay times can be extracted for probing fields $\mathcal{E}_\gamma(t)$ with larger duration $T_{E\gamma} \gg T_{nat}$ than the time scale of the system, simply by tuning the laser frequency through the spectral line width. It is even possible to use an incoherent probe field $\mathcal{E}_\gamma(t)$, where the system interacts at random points in time with photons of various frequencies resolved by a spectrometer. The Fourier transform of the complete linear absorption spectrum corresponds to the response of a system to a (virtual) $\delta$-like excitation event. This is because the response of a system at equilibrium is independent of time, and thus the arrival time of the probing photons is irrelevant. By contrast, the explicitly time-dependent evolution of a non-equilibrium state cannot be accessed by incoherent fields.

The question arises: How can we measure an explicitly time-dependent response?

In the general case of a system undergoing an explicitly time-dependent interaction $V(t)$, the response function reads $d[V(t), t]$ (see Fig.1 for an illustration). Consider the excitation by a coherent laser pulse $\mathcal{E}_\gamma$ with a duration $T_{E\gamma}$ much shorter than the time scale $T_V$ of the perturbation $V(t)$ and $T_{nat}$ of the system ($T_{\mathcal{E}_\gamma} << T_V, T_{nat}$). In this instance the approximation $\mathcal{E}_\gamma(t) = \mathcal{E}_\gamma \cdot \delta(t)$, with $\delta(t)$ the Dirac delta function, can be made and the absorption spectrum from eq. (1) transforms into

$$A(\omega) \propto Im\{\mathcal{F}[d(V(t), t)]\} \qquad \text{eq.(2)}.$$

Due to causality [1, 2, 3], the real part $Re\{\mathcal{F}[d(V(t), t)]\}$, and thus the entire information of the coherent dipole $d[V(t), t]$ excited by the short pulse can be reconstructed. Note that no assumption was made on V(t), making the reconstruction viable even for strongly-nonlinearly driven dynamics. Also, no second (probe) pulse is necessary to sample dynamical information. A single absorption spectrum suffices, without the need for scanning a time delay.

The validity of the reconstruction method is confirmed by first numerically calculating the response $d[V(t), t]$ of a few-level quantum system, which includes a ground state $|0\rangle$ and two excited states $|1\rangle$, $|2\rangle$. They are populated by a weak and short laser pulse, and resonantly coupled by a strong laser field after a fixed time delay (Supplement Section 2). The resulting dipole response, driven by a strong laser pulse, is shown in Fig. 2A. The absorption spectrum $A(\omega)$ is depicted in Fig. 2B. The real-time coherent response of the selected state $|1\rangle$, $d_1[V(t), t]$, is retrieved by using the reconstruction described above. The good agreement between the reconstructed (Fig. 2C) and real (Fig. 2A) coherent dipole response, including the entire time-resolved holographic (amplitude and phase) information, confirms the validity of the approach.

Next, we apply this fully general result to a long-standing problem in the realm of strong-field atomic physics and attosecond science: Strong coupling of autoionizing states [9, 30] in an intense laser field. The experimental setup is displayed in Fig. 3A, employing a typical attosecond transient absorption beamline [5]. With this setup we realize the pulse configuration illustrated in Fig. 1, where first the *2s2p* doubly-excited state of helium is excited by extreme-ultraviolet (XUV) attosecond-pulsed light defining the time $t_0 = 0$ for the measurement. After a fixed time delay, the system interacts with a sub-7-fs (full width at half maximum) near-infrared (NIR) laser pulse (see Fig. 3B for illustration).

The reconstruction method is now applied to the measured line profiles of the *2s2p* state, yielding the time-dependent dipole moment (TDDM). The results are shown in Fig. 4 for a range of intensities. We reconstruct both the amplitude (blue solid lines) and the phase (orange solid lines) of the TDDM $d_{2s2p}(t)$.

The four NIR intensities represent different regimes of NIR strong-field interaction from the weak perturbative regime into the regimes of strong coupling and strong-field ionization of autoionizing states. Fig. 4A,B show the TDDM amplitude with increasingly pronounced minima. As the dipole is directly related to state population, these minima reveal a significant resonant population transfer due to Rabi oscillations between the *2s2p* and *2p²* states. The increasing number of minima with increasing NIR intensity directly follows from the definition of the generalized Rabi frequency $\Omega_R = \sqrt{\Delta^2 + \Omega^2(t)}$, with $\Omega(t) = \mathcal{E}_{NIR}(t) \cdot \mu/\hbar$, and $\mathcal{E}_{NIR}(t)$ being the NIR electric field, $\mu = <2p2|\hat{\mu}|2s2p>$ the transition dipole matrix element connecting both doubly excited states and $\Delta$ the detuning of the laser from the transition frequency. For intensities up to $I_{NIR} = 2.0\ TW/cm^2$ (Fig. 4 A,B) the few-level model (only including $V_{CI}$ (violet/orange dotted lines)) shows good qualitative agreement with the extracted amplitude and phase of the TDDM.

By contrast, for NIR intensities of $I_{NIR} = 6.0\ TW/cm^2$ and up to $I_{NIR} = 10.0\ TW/cm^2$ (Fig. 4 C,D) the experimental results of amplitude and phase greatly differ from the theoretical description that is limited to autoionization and NIR-induced resonant coupling dynamics. The contribution of strong-field ionization pathways, leading to a decrease in the TDDM amplitude, can no longer be neglected. For the given NIR laser parameters of possible ionization into the *N = 2* continuum, the Keldysh parameter $\gamma \approx 2.6 > 1$ suggests that multi-photon ionization should be the favored mechanism [32]. In the absence of known ionization mechanisms for doubly excited states, we employ ionization rates $\Gamma_n = \alpha_n \cdot I_{NIR}(t)^n$, with $I_{NIR}(t)$ being the time-dependent NIR intensity envelope, *n* the order of the process (number of absorbed photons) and $\alpha_n$ are constants that have to be adjusted for each involved state. With this course-grained model the predicted amplitude evolution of the TDDM shows good qualitative agreement with its experimentally reconstructed counterpart, now even for a relatively high NIR intensity of $I_{NIR} = 10.0\ TW/cm^2$. Future development of theory is expected to provide even better agreement, where the full coupling dynamics between all autoionizing states and continua with different symmetries should be included [33]. Interestingly, the experimentally reconstructed phase at such high NIR intensity shows a sign reversal and initially evolves to negative values, which qualitatively differs from the modelled phase evolution of the TDDM. A possible explanation is that at these high NIR intensities, where ionization already plays a major role, the ponderomotive shift of the *2s2p* level, transiently raising its energy [5, 6], is larger than the resonant Stark shift due to coupling to the *2p²* state. The real-time reconstruction of the TDDM presented here thus enables this direct and mechanistically intuitive comparison of theory and experiment in the time domain and the disentangling of different dynamical effects.

The presented method however is far more general, and can be applied to understand complex systems with possibly overlapping resonances or absorption bands in large molecules interacting with strong laser fields [34] in the time domain. The examined superexcited states are an important transient species at the heart of processes as diverse as laser machining, chemical dynamics, plasma dynamics and imaging of molecules at x-ray free-electron lasers (FELs). The real-time reconstruction approach is viable for single-shot (FEL) transmission spectra, where the non-linear response could be used to uncover the *in-situ* timing and ensuing dynamics of x-ray and optical pulses otherwise often lost due to temporal jitter. In the future, this can be utilized to test and develop fundamental quantum theory as well as the measurement of ultrafast non-linear processes in atomic and molecular systems at different time scales. The underlying concept is not limited to the interaction with electric fields and can be more generally applied for the reconstruction of non-equilibrium response functions of any kind of interactions and across all spectral energy regions.

**References:**

[1] R. Kronig, J. Opt. Soc. Amer. **12**, 547 (1926)
[2] H. A. Kramers, Atti. Congr. Fis. **11**, 22 (1927)
[3] R. Kubo, J. Phys. Soc. Japan **12**, 570 (1957)
[4] M. Wu et.al, Phys. Rev. A 88, 043416 (2014)
[5] C. Ott et.al, Science 340, 716-720 (2013)
[6] A. Kaldun, Phys. Rev. Lett. 112, 103001 (2014)
[7] H. Mashiko et.al, Nat. Comm. 5, 5599 (2014)
[8] U. Fano, Phys. Rev. 124, 6 (1961)
[9] P. Lambropoulos, P. Zoller, Phys. Rev. A 24, 379-397 (1981)
[10] I. I. Rabi et.al, Phys. Rev. 55, 526 (1939)

[11] S. R. Leone et.al, Nat.Phot. 8, 162-166 (2014)
[12] R.W. Boyd, Nonlinear optics (Third Edition), 2008
[13] P. Hamm, M. Zanni, Concepts and Methods of 2D Infrared spectroscopy, 2011
[14] E. Goulielmakis et.al, Nature 466, 739-743 (Aug 2010)
[15] H. Wang et.al, Phys. Rev. Lett. 105, 143002 (Oct 2010)
[16] M. Holler et.al, Phys. Rev. Lett. 106, 123601 (Mar 2011)
[17] Z.-H. Loh et.al, Phys. Rev. Lett. 98, 143601 (Apr 2007).
[18] S. Chen et.al, Phys. Rev. A 86, 063408 (2012).
[19] M. Chini et.al, J.Phys. B 47, 124009 (2014).
[20] E. Warrick et.al, J.Phys.Chem. A, 2016, 120 (19), pp 31653174
[21] C.T. Liao, A. Sandhu, *Photonics* **2017**, *4*(1), 17
[22] M. Reduzzi, *JPhys B Vol 49, Nr.6 (2016)*.
[23] M. Schultze et.al, Science 346, 6215, 1348-1352 (2014)
[24] M. Lucchini et.al, Science 353, 6302, 916-919 (2016)
[25] M.Schultze et.al, Nature 493, 75-78 (2013)
[26] A. Assion et.al, Science 282, 5390, 919_922 (1998)
[27] S. Haessler et.al, Phys. Rev. X 4, 021028 (2014)
[28] D. Press et.al, Nature 446, 218-221 (Nov 2008)
[29] J. Mizrahi et. al, Appl. Phys. B (Nov 2013)
[30] H. Bachau, P. Lambropoulos, R. Shakeshaft, Phys. Rev. A 34, 6 (1986)
[31] Z. H. Loh et.al, Chemical Physics 350 (2008) 713
[32] L. V. Keldysh, Sov. Phys. JETP **20**, 1307 (1965)
[33] W. C. Chu, Phys. Rev. A 84, 033426 (2011)
[34] K. Meyer, PNAS, Vol. 112, 51, 15613-15618 (2015)

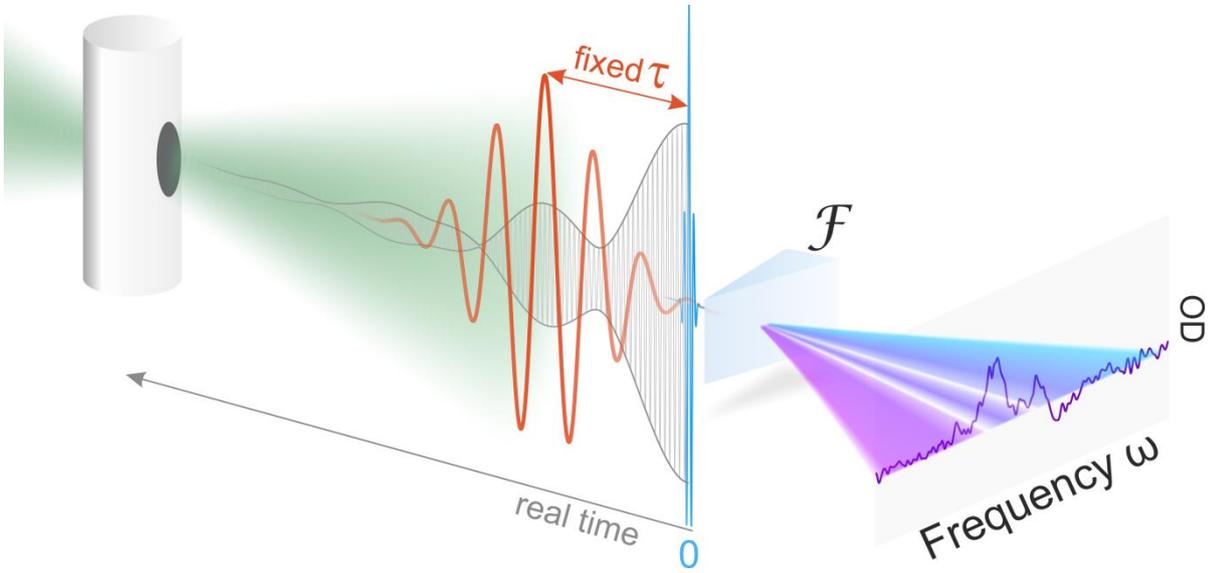

**Figure 1, Non-equilibrium response of matter:** Illustration of the probing of a non-equilibrium state of matter induced by a time-dependent perturbation $V(t)$ using ultra-short laser pulses (blue) to trigger a response (gray), which is then modified e.g. by a strong external time-dependent electric field. From the measured absorption spectrum (eq.(1)) the response can be fully reconstructed if the excitation pulse is much shorter than the system's dynamics (eq.(2)). The interaction with such a $\delta$-like pulse at time 0 produces a causal response ($d[V(t),t] \neq 0$ for $t > 0$ otherwise $d[V(t),t] = 0$) and the Kramers-Kronig relations provide an analytic relation between the real and imaginary part of the Fourier transform of the response (see supplement section 3). Thus, measuring one part is enough to gain access to the full time-dependent information.

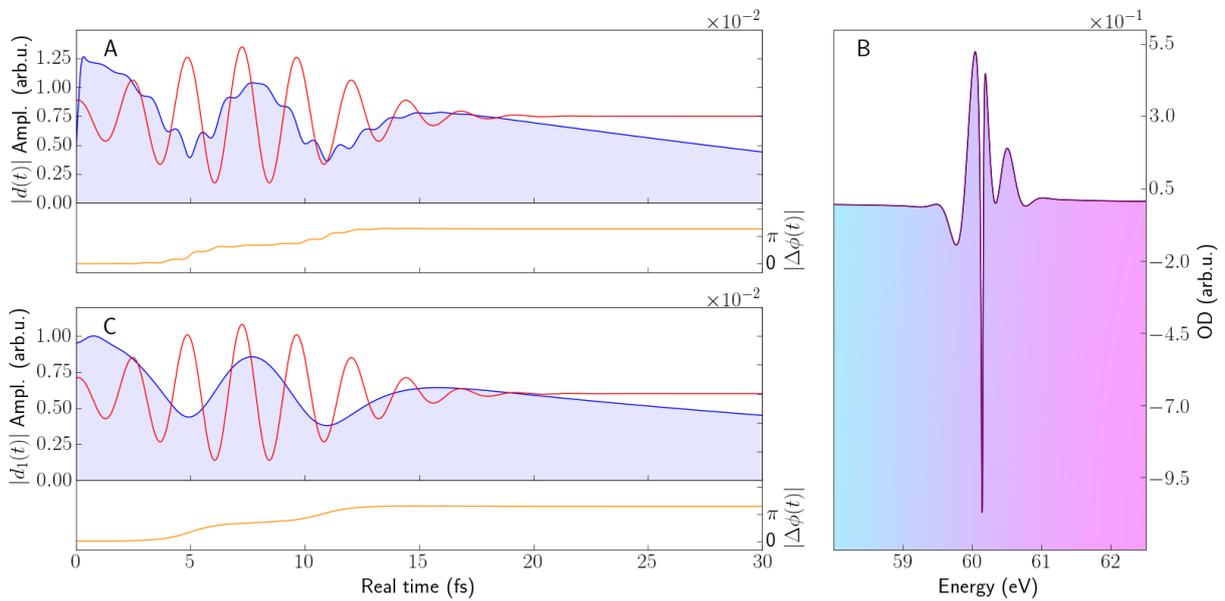

**Figure 2, Reconstruction method:** (**A**) Amplitude (blue curve) and non-trivial phase evolution (orange curve) of the numerically simulated time-dependent response function using a few-level model including two excited states. The carrier frequency is not resolved in this plot. The red curve shows the electric field of the non-linearly interacting laser pulse. (**B**) Calculated optical density from (A) according to eq.(1). (**C**) Test of the reconstruction method of amplitude and phase of the response function by selecting the plotted spectral range and taking the inverse Fourier transform of (B).

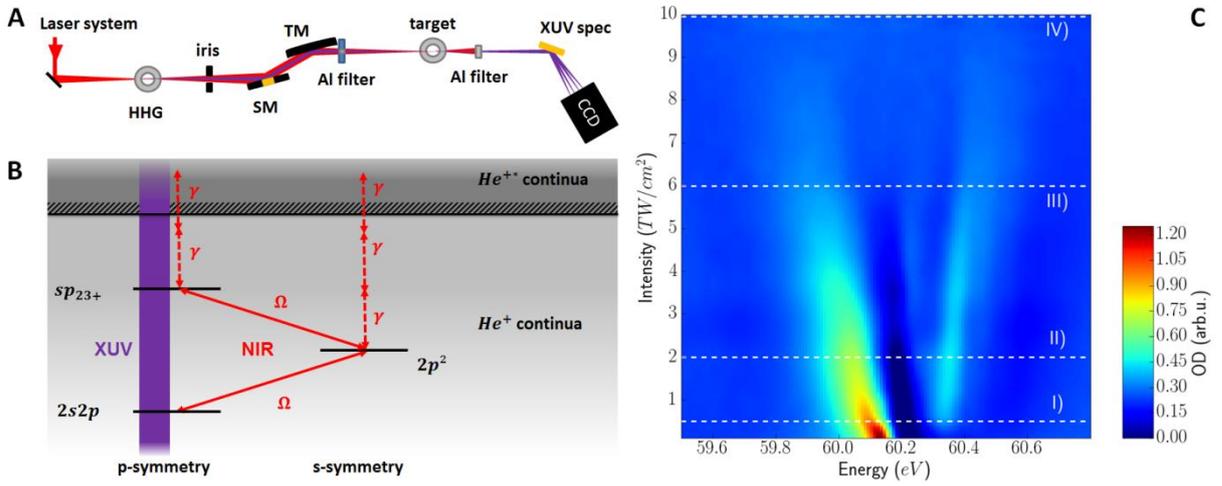

**Figure 3, Experimental setup and data:** (**A**) Illustration of the experimental apparatus. **HHG:** High harmonic generation neon gas target. **SM:** Split-mirror setup for adjusting the XUV-NIR time delay. **TM:** Toroidal mirror. **XUV spec:** XUV flat field spectrometer (spectral resolution: 50 meV FWHM). (**B**) Level scheme of the relevant doubly excited states of helium. The NIR pulse induces resonant coupling $\Omega$ between the autoionizing doubly excited states and also leads to resonantly enhanced strong-field multi-photon ionization $\gamma$ into the $N=2$ continuum at the highest intensities of up to 10 TW/cm$^2$. The included pathways are (**C**) NIR intensity scan of Autler-Townes (AT) splitting [31] in doubly excited helium around the *2s2p*-state at 60.15 eV at fixed time delay $\tau$=7.4±0.1fs. The high spectrometer resolution allows the observation of the line shape in great detail. White dashed lines (I)-(IV) indicate the spectra used in the reconstruction discussed in Fig. 4.

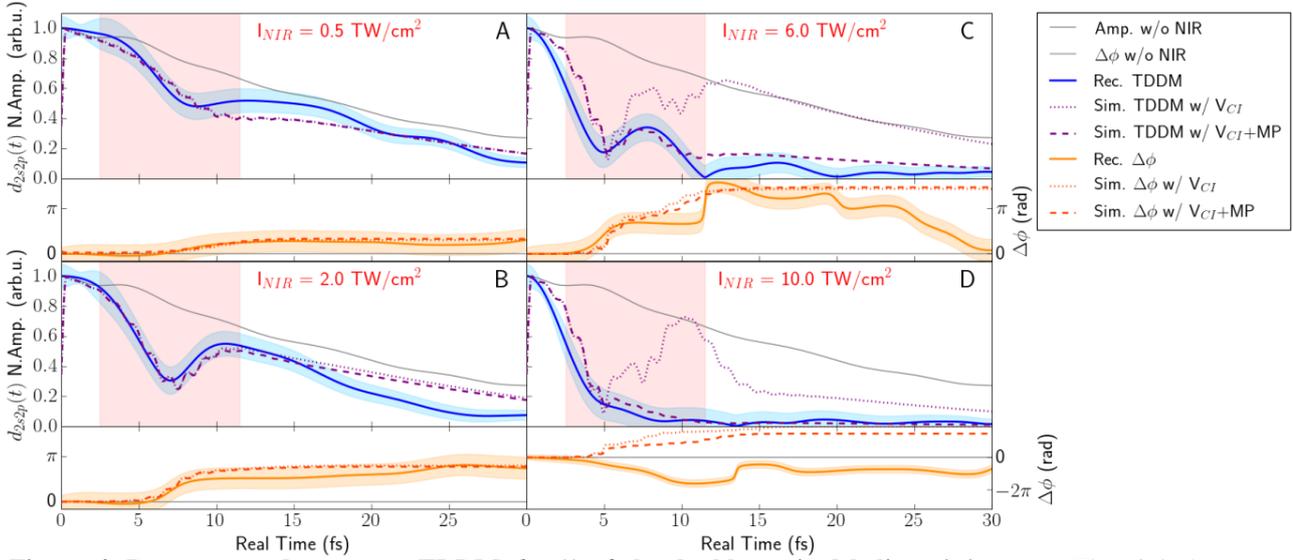

**Figure 4, Reconstructed response, TDDM $d_{2s2p}(t)$ of the doubly excited helium 2s2p state:** The violet/orange curves show the numerically calculated amplitude/phase evolution of the TDDMs including only configuration interaction $V_{CI}$ pathways (dotted), and additionally, a multi-photon ionization model (dashed), which becomes increasingly important at the highest laser intensities. The gray lines show the TDDM for zero NIR intensity with the expected decay due to autoionization. The shaded blue area represents the $1\sigma$-error of the reconstructed TDDM (see supplement). The shaded red area indicates the time where the NIR pulse interacts with the system. **(A)-(B)** Reconstructed TDDM amplitude and phase showing the emerging minima at peak NIR intensity indicative of Rabi oscillations due to resonant coupling between the *2s2p* and *2p²*-state for the intensities $I_{NIR}$ = 0.5, 2.0 TW/cm². **(C)-(D)** The TDDM becomes strongly modulated with shifted and additional minima. The onset of strong-field ionization and the resulting depletion of the states during the NIR pulse for even higher intensities at $I_{NIR}$ = 6.0, 10.0 TW/cm² is visible.